\documentclass[10pt,aps,pre,twocolumn,nopacs,final,letterpaper,superscriptaddress,longbibliography]{revtex4-2}
\usepackage[utf8]{inputenc}
\usepackage{amsmath,amsfonts,amssymb}
\usepackage{graphicx}
\usepackage{bm}
\usepackage{hyperref}
\usepackage{xcolor}
\usepackage{quotes}
\definecolor{goodblue}{RGB}{0, 91, 187}
\hypersetup{
  colorlinks=true,
  allcolors=goodblue,
  urlcolor=goodblue,
  citecolor=goodblue,
  pdfborder={0 0 0},
  breaklinks=true,
}

\newcommand{\boldA}{\bm{A}}
\newcommand{\boldQ}{\bm{Q}}
\newcommand{\boldX}{\bm{X}}
\newcommand{\bbN}{\mathbb{N}}
\newcommand{\boldx}{\bm{x}}
\newcommand{\boldc}{\bm{c}}

\newcommand{\boldone}{\bm{1}}
\newcommand{\ER}{Erd\H os-R\'enyi }

\begin{document}

\title{Reconstructing networks from simple and complex contagions}
\author{Nicholas W. Landry}
\email{nicholas.landry@virginia.edu}
\affiliation{Vermont Complex Systems Center, University of Vermont, Burlington, Vermont 05405, USA}
\affiliation{Department of Mathematics and Statistics, University of Vermont, Burlington, Vermont 05405, USA}
\affiliation{Department of Biology, University of Virginia, Charlottesville, Virginia 22903, USA}
\author{William Thompson}
\email{william.thompson.2@uvm.edu}
\affiliation{Vermont Complex Systems Center, University of Vermont, Burlington, Vermont 05405, USA}
\author{Laurent H\'ebert-Dufresne}
\email{laurent.hebert-dufresne@uvm.edu}
\affiliation{Vermont Complex Systems Center, University of Vermont, Burlington, Vermont 05405, USA}
\affiliation{Department of Computer Science, University of Vermont, Burlington, Vermont 05405, USA}
\author{Jean-Gabriel Young}
\email{jean-gabriel.young@uvm.edu}
\affiliation{Vermont Complex Systems Center, University of Vermont, Burlington, Vermont 05405, USA}
\affiliation{Department of Mathematics and Statistics, University of Vermont, Burlington, Vermont 05405, USA}

\begin{abstract}
Network scientists often use complex dynamic processes to describe network contagions, but tools for fitting contagion models typically assume simple dynamics. Here, we address this gap by developing a nonparametric method to reconstruct a network and dynamics from a series of node states, using a model that breaks the dichotomy between simple pairwise and complex neighborhood-based contagions. We then show that a network is more easily reconstructed when observed through the lens of complex contagions if it is dense or the dynamic saturates, and that simple contagions are better otherwise.
\end{abstract}

\maketitle

Contacts between individuals are how diseases, information, and transmissible social behavior spread through a population.
Classical epidemic models treat these contact as well-mixed \cite{kermack_contribution_1927}, but we now have overwhelming evidence that the structure of these social systems can strongly influence the outcomes of a contagion---for example, the epidemic threshold~\cite{pastor-satorras_epidemic_2001}, extent \cite{miller_note_2012,hebert-dufresne_r0_2020} or the time-scale of spread and the distribution of secondary infections \cite{althouse_superspreading_2020}.  
Despite the strength of the evidence for structural models, network methods remain difficult to deploy beyond the theoretical laboratory, partly because detailed contact patterns are seldom observed and instead need to be inferred from messy and incomplete observational data \cite{peel_statistical_2022,young_bayesian_2021}.
In fact, this inference problem---network reconstruction---is still a fundamental problem in network science~\cite{peel_statistical_2022,shandilya_inferring_2011}.

Recent progress suggests that state-of-the-art reconstruction is possible with Bayesian frameworks~\cite{peixoto_network_2019,gray_bayesian_2020} in cases where the contagion spreads via single exposures to infectious individuals~\cite{peixoto_network_2019}.
Other recent work has shown that network reconstruction is possible using measurements from repeated simple cascades~\cite{gray_bayesian_2020}, or binary state time series \cite{ma_statistical_2018}.
Less is known about the inference of social contagions, often thought to spread due to multiple exposures \cite{valente_network_1996,watts_influentials_2007,centola_spread_2010}.
Previous studies have focused on distinguishing different types of contagions, whether interacting contagions and social reinforcement \cite{hebert-dufresne_macroscopic_2020}, simple and complex contagions \cite{murphy_deep_2021,cencetti_distinguishing_2023}, or heterogeneous and complex contagions \cite{st-onge_nonlinear_2024}, but these assume the underlying contact network is known and there is no emphasis on reconstruction itself.

In this Letter, we develop a nonparametric approach to network reconstruction from contagion data that does not assume the contagion dynamics are simple or complex.
Using a time series of binary node states as input, we show how to jointly identify the network's structure and the dynamical rules that generated the contagion.
Our method is conceptually related to that of Refs.~\cite{liu_inferring_2021,ma_statistical_2018,liu_expectationmaximizing_2023}, which all reconstruct a network and its dynamical rules from binary time-series.
Unlike these previous studies, however, we put few restrictions on the contagion rules, and we generate a complete posterior distribution over parameters and contagion rules instead of point estimates.\\
We then use this framework to examine the difficulty of network reconstruction as a function of the contagion process's rules.\\

\paragraph*{\textbf{Neighborhood-based contagions on networks.}} 
We describe network contagions with a neighborhood-based susceptible--infected--susceptible (SIS) model~\cite{dodds_universal_2004}, which encapsulates several simple and complex contagion processes as special cases.
This contagion process is defined as follows: At time $t$, each infected nodes recover with probability $\gamma$ and each susceptible nodes become infected with probability $c(\nu, \theta)$, where $c:\bbN\mapsto [0, 1]$ is the \textit{contagion function}, $\nu$ is the number of infected neighbors of a node, and $\theta$ denotes any function parameters. 
We write this function as a contagion vector $\boldc = [c_0, \dots, c_{N-1}]^T$  in nonparametric form, where $c_\nu$ denotes the probability of infection by $\nu$ infected neighbors and $N$ is the number of nodes.

The classical network SIS and threshold models are special cases of this general model~\cite{dodds_universal_2004}.
In discrete simple contagions, a node is infected with probability $\beta$ from a single exposure, and exposures are modeled as independent, meaning that the contagion function is $c(\nu, \beta) = 1 - (1 - \beta)^\nu$.
In complex contagions, multiple exposures are required for the process to spread, and they are most commonly studied using the threshold model \cite{valente_network_1996,watts_influentials_2007}, where a node adopts an opinion when the number of its neighbors holding that opinion exceeds a critical threshold $\tau$. 
In that case, the contagion function becomes $c(\nu, \beta, \tau) = \beta \boldone_{\nu\geq \tau}$, where $\beta = 1$. 
This nonparametric model allows us to define a reconstruction process unrestricted by these two particular cases.

To describe the time evolution of this dynamical process mathematically, we denote the states of all nodes with a vector  $\boldx(t)=[x_1(t),\,\dots,\,x_N(t)]^T$, where $x_i(t)$ is the infection status of individual $i$ at time $t$, with $x_{i}(t)=0$ corresponding to a susceptible state and $x_{i}(t)=1$ to the infected state.
We track the states of nodes at unit time intervals $t = 0, 1,\dots, T$. 
The collection of the state vectors is then the matrix $\boldX = [\boldx(0),\,\boldx(1),\dots,\, \boldx(T)]$.

\begin{figure}
  \centering
  \includegraphics[width=8.6cm]{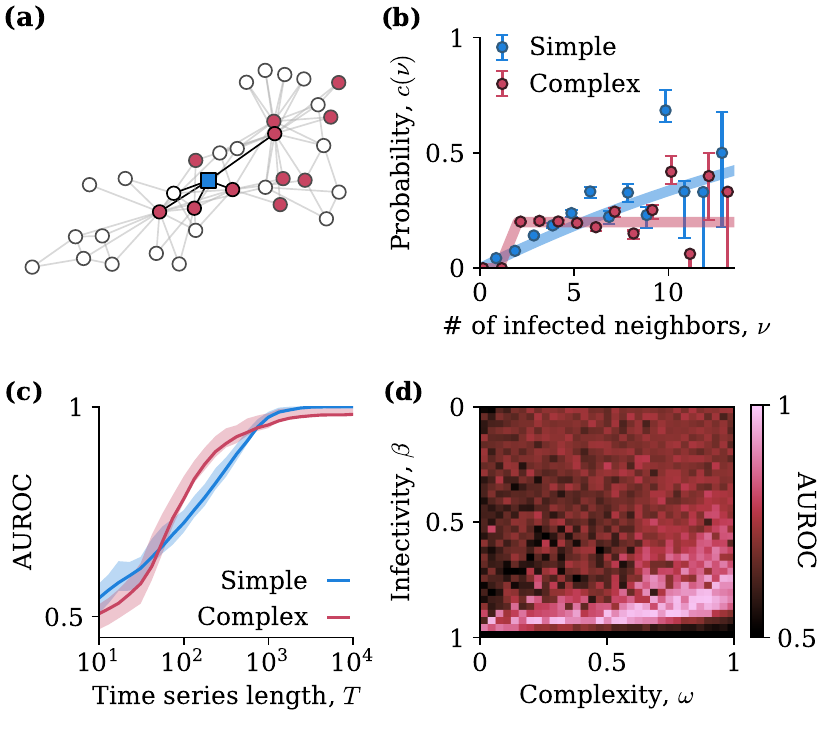}
  \caption{
    \label{fig:framework} Overview of the problem.
    \textbf{(a)} 
    A contagion spreads on a network for $T$ time steps, and we observe the resulting sequence of states $\boldX$.
    The probability that a susceptible node (white) becomes infected (red) at the next time step is a function $c(\nu)$ of the number of infected neighbors it has, e.g., $\nu=4$ for the square node highlighted in blue.
    \textbf{(b)} 
    We compute a nonparametric Bayesian estimate of the contagion function $c(\nu)$. 
    Here, we show an estimate of $c(\nu)$ obtained from a single short realization of the dynamics when the network is known. Error bars show the 50\% highest-density posterior interval (HDPI) of $c(\nu)$.
    \textbf{(c)} 
    We estimate the network and the contagion function $c(\nu)$ simultaneously using the marginals of the posterior distribution, Eq.~\eqref{eq:joint_posterior}. 
    The reconstruction error goes to 0 as the amount of data $T$ goes to infinity. 
    The shaded regions indicate the 50\% HDPI, and lines show the median AUROC across $10^3$ repetitions.
    \textbf{(d)} 
    The reconstruction quality is determined by the shape of the contagion function, here demonstrated by varying its overall infectivity $\beta$ and the level of complexity $\omega\in[0,1]$.
    We use the parametrization $c(\nu,\beta,\omega) = (1 - \omega) g + \omega h$ where $g(\nu, \beta) =  1 - (1-\beta)^{\nu}$ describes a simple contagion model, and $h(\nu, \beta) =\beta \boldone_{\nu \geq 2}$ describes a complex threshold model.
  }
\end{figure}

All the transition probabilities of the dynamics can be compactly expressed as 
\begin{multline}
  P\left(x_i(t+1) \mid x_i(t),\, \boldA,\, \theta\right)=\\
  \qquad b\bigl( \gamma,\, 1 - x_i(t+1)\bigr)^{x_i(t)}\  b\bigl(c_{\nu_i(t)},\, x_i(t+1)\bigr)^{1 - x_i(t)},
  \label{eq:bernoulli_transitions}
\end{multline}
where $\theta$ denotes all model parameters collectively---the recovery rate and the nonparametric contagion functions---and
where $b(p,\, x) = p^x (1 - p)^{1-x}$ is the Bernoulli probability mass function.
In the absence of additional dependencies, we can write the joint probability of the transitions from step $t$ to step $t+1$ as 
\begin{multline}
  P(\boldx(t+1) \mid \boldx(t),\, \boldA,\, \theta) \\= \prod_{i \in V} P(x_i(t+1) \mid x_i(t),\, \boldA,\, \theta).
  \label{eq:state_independence}
\end{multline}
The joint probability of a sequence of states $\boldX$ is then given as the product
\begin{align}
  P(\boldX = \boldx \mid \boldA,\, \theta)  = \prod_{t=1}^{T - 1} P\left(\boldx(t+1) \mid \boldx(t),\, \boldA,\, \theta\right),
  \label{eq:markovian}
\end{align}
since the dynamics are Markovian.
By substituting Eq.~\eqref{eq:bernoulli_transitions} in Eq.~\eqref{eq:state_independence} and then in Eq.~\eqref{eq:markovian} and re-arranging terms, we can finally rewrite this probability as
\begin{align}
&P(\boldX \mid \boldA, \theta)= f(\gamma, \boldX)\prod_{\ell=0}^{N-1}c_{\ell}^{m_{\ell}}  (1 - c_{\ell})^{n_{\ell}},
\label{eq:X_cond_A_c}
\end{align}
where
\begin{subequations}
\begin{align}
m_{\ell} &= \sum_{t=0}^{T-1}\sum_{i \in V} [1 - x_i(t)]x_i(t+1)\boldone_{\nu_i(t) = \ell},\\
n_{\ell} &= \sum_{t=0}^{T-1}\sum_{i \in V} [1 - x_i(t)][1 - x_i(t+1)]\boldone_{\nu_i(t) = \ell},
\end{align}
\end{subequations}
are the counts of infection events occurring and failing to occur when the number of infected neighbors is $\ell$, and where $f(\gamma, \boldX)$ is a function of $\gamma$  and $\boldX$, but not of $\boldA$.
(This factorization is due to Eq.~\eqref{eq:bernoulli_transitions}, where  $\nu_i(t)$ is the only term that depends on $\boldA$, implicitly.)

These equations determine the probability of a particular sequence of node states occurring for a given adjacency matrix $\boldA$ and set of dynamical parameters $\theta$ and, through Bayesian inference, they can be reversed to obtain the probability of these structural and dynamical properties given a sequence of system states.
Figure~\ref{fig:framework} highlights this framework, which we now describe in more detail.\\

\paragraph*{\textbf{Network and dynamic reconstruction.}}
Applying Bayes' rule, we write
\begin{equation}
P(\boldA, \theta,\rho \mid \boldX) =\\
  \frac{P(\boldX \mid \boldA, \theta)\ P(\boldA \mid \rho)\ P(\rho, \theta)}{P(\boldX)},
\end{equation}
where $P(\boldA \mid \rho)$ is our prior over network structures $\boldA$, and where $\rho\in[0,1]$ is a parameter of the network model, a simple \ER model of expected density $\rho$.
This model assigns a probability $P(\boldA \mid \rho) = \rho^{|E|}(1 - \rho)^{\binom{N}{2} - |E|}$  to networks with $|E|$ edges on $N$ nodes.
(But see Ref.~\cite{peixoto_network_2019} on replacing this simplistic assumption with more structured models.)
Noticing that the parameters of the dynamical process $\theta=(\gamma, \boldc)$ and the density $\rho$ are all defined on the interval $[0,1]$, we choose independent (conjugate) beta priors with hyperparameters $(a_{\gamma}, b_{\gamma})$ for $\gamma$, and likewise for all the other parameters.

With these modeling choices, the posterior distribution can be written in closed form as
\begin{align}
  P(\boldA, \theta,\rho \mid \boldX) \propto\
  & f(\gamma, \boldX)\times \prod_{\ell=0}^{N-1}c_{\ell}^{m_{\ell} + a_\ell - 1}  (1 - c_{\ell})^{n_{\ell} + b_\ell - 1} \notag\\
  &\times  \rho^{|E| + a_\rho - 1}(1 - \rho)^{\binom{N}{2} - |E| + b_\rho - 1},
  \label{eq:joint_posterior}
\end{align}
where we have dropped normalization terms and redefined $f(\gamma,\boldX)$ to include the prior density of $\gamma$ also.

The network structure $\boldA$ is the primary target of network reconstruction, so we first focus on the marginal posterior distribution, which is obtained by integrating Eq.~\eqref{eq:joint_posterior} over contagion vectors $\boldc$, recovery rates $\gamma$, and densities $\rho$. This yield
\begin{align}
P(\boldA \mid \boldX) 
  &\propto \mathcal{B}\left(|E|+a_{\rho},\binom{N}{2}-|E|+b_{\rho}\right)\nonumber\\
  &\quad\times\prod_{\ell = 0}^{N-1}\mathcal{B}(m_{\ell}+a_{\ell},n_{\ell}+b_{\ell}),
\end{align}
where $\mathcal{B}(a, b) = \Gamma(a)\Gamma(b)/\Gamma(a+b)$ is the beta function.
We generate samples from this distribution with a simple edge-flip Markov chain Monte Carlo (MCMC) algorithm~\cite{peixoto_network_2019}.
We can then use these samples to compute a matrix $\boldQ$ of edge probability estimates, $\boldQ = \sum_{s = 1}^{N_s} A^{(s)} /N_s$  where $N_s$ is the number of samples.
All our results are computed using chains starting from random initial conditions, with samples taken every $10^4$ steps following an initial burn-in of $10^5$ steps.

We augment these network samples with estimates of the dynamical parameters, $\theta=(\gamma, \boldc)$, using the relation
\begin{align*}
P(\gamma,\, \boldc, \rho \mid \boldX,\, \boldA)  \propto P(\gamma \mid \boldX,\, \boldA)P(\boldc \mid \boldX,\, \boldA)P(\rho \mid \boldX,\, \boldA),
\end{align*}
which is obtained by conditioning Eq.~\eqref{eq:joint_posterior} on $\boldA$, dropping all constants and noticing that the results factorize as a product of beta densities.
More specifically, we find that $P(\gamma \mid \boldX,\, \boldA)$ is a beta distribution with parameters $(h + a_\gamma, g + b_\gamma)$, where
\begin{align*}
g &= \sum_{t = 0}^{T-1} \sum_{i \in V} x_i(t) x_i(t+1),\\
h &= \sum_{t = 0}^{T-1} \sum_{i \in V}  x_i(t)[1 - x_i(t+1)].
\end{align*}
Similarly, the nonparametric infection probabilities $\boldc$ follow beta distributions $\boldc_\ell \sim \mathrm{Beta}(m_{\ell} + a_{\ell}, n_{\ell} + b_{\ell})$. There is no need to sample $\rho$ directly as $\boldA$ offers a good summary of structure; nonetheless, if needed, its posterior density is also a beta distribution.

\paragraph*{\textbf{Results.}}
Figure~\ref{fig:framework} illustrates our nonparametric Bayesian approach using the Zachary's Karate Club network (henceforth ZKC) \cite{zachary_information_1977} as an example.
We generate synthetic time series data comprising nodal states for $T=10^3$ time steps with a recovery rate of $\gamma = 0.1$ starting from the initial condition $\boldx(0)=\boldone$ to prevent premature stochastic extinctions.
We study the impact of the contagion function by running these simulations with both the SIS model, defined by $c(\nu, \beta)= 1 - (1 - \beta)^\nu$, and the threshold contagion model $c(\nu, \beta, \tau)= \beta \boldone{\nu\geq \tau}$ with $\tau=2$ (we also considered $\tau = 3$ in the Supplemental Material).
We find that imperfect reconstruction of the contagion function $\bm{c}$ is possible with short time series, as illustrated in Fig.~\ref{fig:framework}(b).
Values of $\nu$ for which we have little data---corresponding to high degrees---are more difficult to estimate accurately, as our nonparametric framework does not benefit from the strong inductive biases of simpler models.

Figure~\ref{fig:framework}(c) describes the quality of the network reconstruction for ZKC, as measured by the Area Under the Receiver Operating Characteristic (AUROC)  \cite{hanley_meaning_1982},  as a function of the amount of time-series data.
We calculate the AUROC by treating the edges $\boldA$ of the ZKC as binary labels and edge probabilities $\boldQ$ as posterior estimates for these labels.
To make this comparison meaningful, we first select $\beta=0.04$ for the simple contagion process and then match the maximum number of infection events per node of both processes in expectation by tuning the value $\beta$ of the complex contagion process with the Robbins-Monro algorithm \cite{robbins_stochastic_1951}. 
Both processes thus generate a similar \emph{amount} of transition data, but their quality may vary.

We observe an intermediate value of $T$ where complex contagions reconstruct the network more accurately than simple contagions.
In addition, we find that increasing the amount of data leads to diminishing returns in reconstruction accuracy and that the accuracy reaches a plateau close to $1$ for both dynamics in the limit of large $T$.
Lastly, we construct a mixture of contagion functions parametrized by $\omega\in[0,1]$ (where $\omega=0$ corresponds to a simple contagion and $\omega=1$ to a threshold contagion)
Fig.~\ref{fig:framework}(d) demonstrates that network reconstruction depends on both the infectivity of the contagion function and the complexity of the contagion.

\begin{figure}
  \centering
  \includegraphics[width=8.6cm]{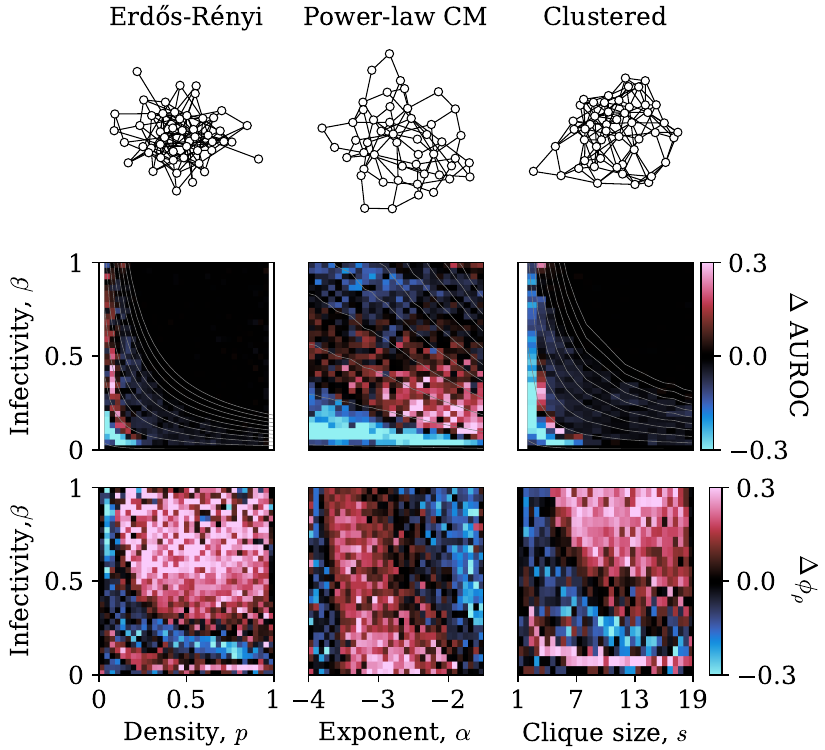}
  \caption{\label{fig:compare_sis_tau2}
  The average difference in reconstruction performance between simple and complex contagions for the SIS model and the threshold model with $\tau=2$.
  Red denotes regions where complex contagions outperform simple contagions.
  Blue denotes regions where simple contagions outperform complex contagions.
  The first row visualizes a sample from each network model; the second row compares the performance of the network reconstruction; and the third row compares the estimation of the network density.
  The gray lines show the basic reproduction number $R_0=\beta\sigma(\boldA)/\gamma\in\{1,11,21,...\}$ \cite{diekmann_definition_1990}, where $\sigma(\boldA)$ is the spectral radius of $\boldA$, and grow when moving from the lower left corner to the upper right.
  }
\end{figure}

Having confirmed that the method works well, we then investigate whether complex or simple contagion models lead to an easier network reconstruction problem. 
We explore this question through a comprehensive simulation experiment with generative network models that capture structural properties observed in empirical networks.
The \ER model \cite{erdos_random_1959}, which we use to examine the effect of network density, is characterized by $p$, the probability of two nodes connecting at random.
We use the network configuration model (CM) \cite{fosdick_configuring_2018} with a degree distribution $p(k)\propto k^{\alpha}$ supported on $[2, N-1]$ to observe the effects of degree heterogeneity, and we comment that when $\alpha$ is increased, the network density also increases.
The clustered network model \cite{newman_properties_2003}, which we use to represent clique structure, is constructed by forming a bipartite network, where type-1 nodes have degree 2 and type-2 nodes have degree $s$, and then projecting that network onto the type-1 nodes to create cliques of size $s$.
The mean degree of the projected network scales with the square of the clique size. 
We fix the network size to $N=50$ nodes and generated time series of $T = 2~\times~10^3$ steps.
(We ran the study on two additional models; see the Supplemental Material for more details.)

Figure~\ref{fig:compare_sis_tau2} summarizes the result of this experiment using the \emph{difference} in AUROC, with positive values (red) corresponding to regions of the parameter space where complex contagions outperform simple contagions.
To complement this information, we also calculate the network density estimate quality, $\phi_\rho = 1 - \sum_{s=1}^{N_s}|\hat{\rho}_s - \rho|/N_s$ and compute a difference in performance  $\Delta \phi_\rho$.

We find structural and dynamical regimes where different types of contagions outperform each other.
The regions where simple contagions outperform complex contagions can be sufficiently explained by the basic reproduction number $R_0$, calculated as $\beta\sigma(\boldA)/\gamma$, where $\sigma(\boldA)$ is the spectral radius of $\boldA$.
For all models, the simple contagion process outperforms the complex contagion process in a region roughly corresponding to $2\leq R_0\leq 6$ in the simple contagion.
For higher $R_0$ values, the complex contagion process outperforms simple contagion as the infection time series saturates.
Complex contagion never generates transitions with $\nu=1$ since the threshold forbids it; this allows the algorithm to infer denser substructures without noise introduced by pairwise infection information.
Just above this region, there is another region for large enough $R_0$ where simple contagion again outperforms complex contagion.
This can be understood because complex contagion infers a network close to a complete network, whereas simple contagion infers an empty network---but with high confidence about several of the network links.
For the densest of networks, these algorithms perform equally (notice that the densest power law CM is far from a complete network), because there is so much saturation that the elimination of pairwise information is not enough to overcome the confounding noise present when almost every node is infected at once.

There may be cases where, although our algorithm fails to place links correctly, it nonetheless can accurately recover the \textit{number} of links, i.e., the density, $\rho$.
Comparing the performance of complex contagions with respect to simple contagions in the bottom row of Fig.~\ref{fig:compare_sis_tau2}, we observe that these regions do not correspond to the performance differences with respect to AUROC.
The regions in the upper right of the plots for the \ER and clustered network models where complex contagions vastly outperform simple contagions in estimating the network density are caused by bimodality in the distribution of estimated $\rho$ values inferred from simple contagion time series. 
This is because our MCMC sampler struggles to sample in this region and gets stuck in local maxima due to the ruggedness of the likelihood landscape.
In the lower left of the plots, where the variance in the estimated value of $\rho$ is lower, we see similar behavior as described above for the AUROC.

\begin{figure}[b]
  \centering
  \includegraphics[width=6cm, trim=0.5cm 0.5cm 0cm 0.9cm, clip=true]{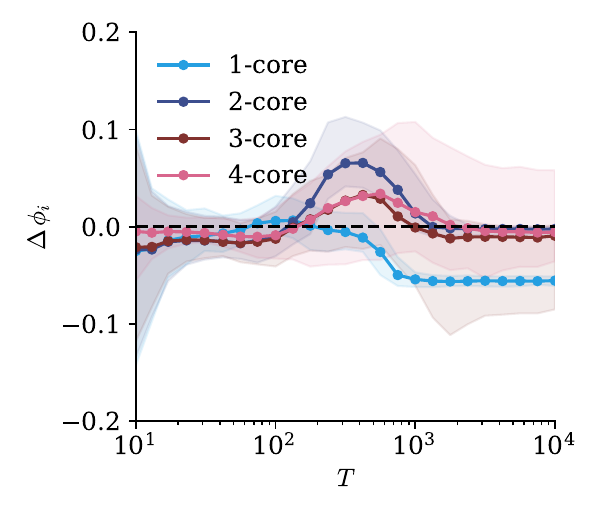}
  \caption{\label{fig:nodal_performance}
  Difference in reconstruction performance of the network SIS contagion process and the threshold contagion process with threshold $\tau=2$, for nodes of various coreness and dynamical parameters identical to those chosen in Fig.~\ref{fig:framework}(c).
  This illustrates that for intermediate amounts of infection data, complex contagions outperform simple contagions due to their recovery of $k \geq 2$-core nodes.
  The shaded regions represent the 50\% HDPI, and the lines are the median of $10^3$ realizations.
  }
\end{figure}

A distinguishing feature of the threshold contagion is that it cannot spread to nodes within the 1-core.
We observe this in Fig.~\ref{fig:nodal_performance}, where we revisit the ZKC results to study the reconstruction performance of nodes with specific coreness values, calculated as the average of
\begin{equation}
 \phi_i = 1 -\frac{1}{N_s}\sum_{s = 1}^{N_s}\left(\frac{1}{N}\sum_{j = 1}^{N} |\boldQ_{ij} - \boldA_{ij}| \right)
\end{equation}
over all nodes $i$ within the same coreness class.
We find that nodes of coreness greater than one are responsible for complex contagions outperforming simple contagions in the region before AUROC plateaus [see Fig.~\ref{fig:framework}(c)].
Similarly, as the amount of data is increased with $T$, we see that the differential performance of 1-core nodes seems to drive the improvement in the performance of simple contagions compared with complex contagions.

\paragraph*{\textbf{Discussion.}}
By developing a nonparametric approach to the network inference problem, we avoid model-based biases and are able to study how the complexity of a dynamic process helps or hurts our ability to infer its rules and the network that supports it.
We find that complex contagions can have an advantage over equivalent simple contagions if they can avoid saturation and better resolve dense networks.

This result suggests that different dynamical processes have different statistical power to resolve different network structures, with threshold contagions being better in dense cliques when the dynamics are sufficiently intense.
More generally, we could imagine optimizing or tuning a dynamic process to infer different network features.
Future work should, therefore, explore other network structures and other types of contagions (e.g., relative thresholds, superlinear contagions). In doing so, we should be able to categorize different types of contagions not just by their global accuracy but by how well they can resolve different local features of networks. 

More practically, it also suggests that different data streams allow us to learn different types of network features depending on the complexity of their generative mechanisms.
Our results show that network reconstruction is much more feasible for diseases such as SARS-CoV-2 \cite{li_genetics_2023}, Mpox \cite{guzzetta_early_2022}, or rhinovirus in contrast to highly infectious diseases such as measles \cite{guerra_basic_2017} or chickenpox \cite{lee_estimating_2021}.
In the context of social media where the spread of information is often modeled as a complex contagion process, our results imply that we will learn less than would be expected if we simply translate prior reconstructability results for simple contagion processes \cite{peixoto_network_2019} with empirically determined estimates of $R_0$ \cite{kucharski_modelling_2016}.
For extremely viral trends or information, however, we may be able to recover the effective contact networks with more precision than previously estimated.
These results will not only inspire future network reconstruction methods but also guide how well different types of experimental data can hope to inform network reconstruction.

\paragraph*{\textbf{Data availability.}} All code and data used in this study are available on \href{https://github.com/nwlandry/complex-network-reconstruction}{Github} and at Ref.~\cite{landry_code_2024}. All networks are visualized with XGI \cite{landry_xgi_2023}.

\paragraph*{\textbf{Funding.}}
N.W.L., L.H.-D., and J.-G.Y. acknowledge support from the National Institutes of Health 1P20 GM125498-01 Centers of Biomedical Research Excellence Award. N.W.L. acknowledges support from the University of Virginia Prominence-to-Preeminence (P2PE) STEM Targeted Initiatives Fund, SIF176A Contagion Science.

\bibliography{references}

\clearpage
\onecolumngrid
\section{Supplemental Material}

\renewcommand{\thefigure}{S\arabic{figure}}
\setcounter{figure}{0}
\renewcommand{\thetable}{S\arabic{table}}
\setcounter{table}{0}
\renewcommand{\theequation}{S\arabic{equation}}
\setcounter{equation}{0}

\paragraph*{\textbf{Additional models.}}
We also considered network models with constant density, specifically, the small-world network model and the stochastic block model (SBM) \cite{holland_stochastic_1983}.
The small-world network model \cite{watts_collective_1998}, which we use to study the effect of "small-worldness" and the clustering coefficient is constructed with a ring network composed of degree 6 nodes and rewired these links with probability $p$.
The SBM, which we use to examine community structure, was constructed with two communities of equal size and a constant mean degree of $10$.
The \textit{imbalance parameter} \cite{jerrum_metropolis_1998}, $\epsilon$, interpolates between the two extreme cases of an \ER network without community structure ($\epsilon=0$) and two disconnected communities ($\epsilon=1$).

\paragraph*{\textbf{Additional results.}}
In this section, we present additional results supporting our conclusions in the main text. Firstly, we compare the performance of simple and complex contagion, but now, where complex contagion is a threshold model with $\tau=3$.

\begin{figure}
    \centering
    \includegraphics[width=8.6cm]{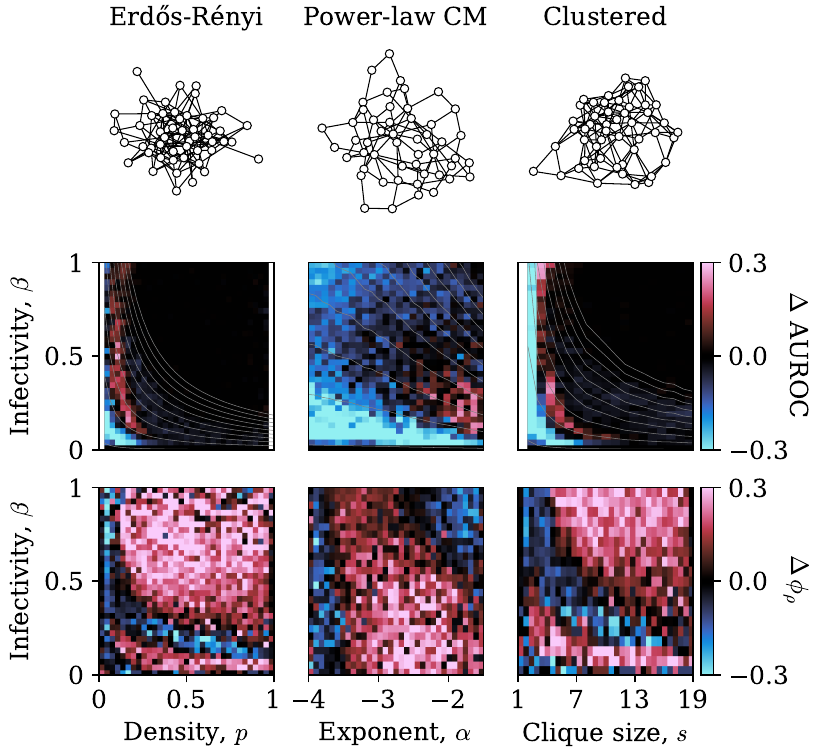}
    \caption{\label{fig:compare_sis_tau3}
  The average difference in reconstruction performance between simple and complex contagion for the SIS model and the threshold model with $\tau=3$.
  Red denotes regions where complex contagions outperform simple contagions.
  Blue denotes regions where simple contagions outperform complex contagions.
  The first row visualizes a sample from each network model; the second row compares the performance of the network reconstruction; and the third row compares the estimation of the network density.
  The gray lines show the basic reproduction number $R_0=\beta\sigma(\boldA)/\gamma\in\{1,11,21,...\}$, and grow when moving from the lower left corner to the upper right.
    }
\end{figure}

Fig.~\ref{fig:compare_sis_tau3} illustrates much of the same behavior as seen in Fig.~\ref{fig:compare_sis_tau2}, with the exception of the AUROC for the CM.
This can be understood because the minimum degree of the power-law degree sequence is 2.
When the exponent of the power law is small enough, the degree sequence is quite homogeneous, and the network can be seen as a perturbation of a 2-regular network, which is insufficiently connected to sustain a contagion that requires three infected neighbors to infect a node.

\begin{figure*}
    \centering
    \includegraphics[width=\textwidth]{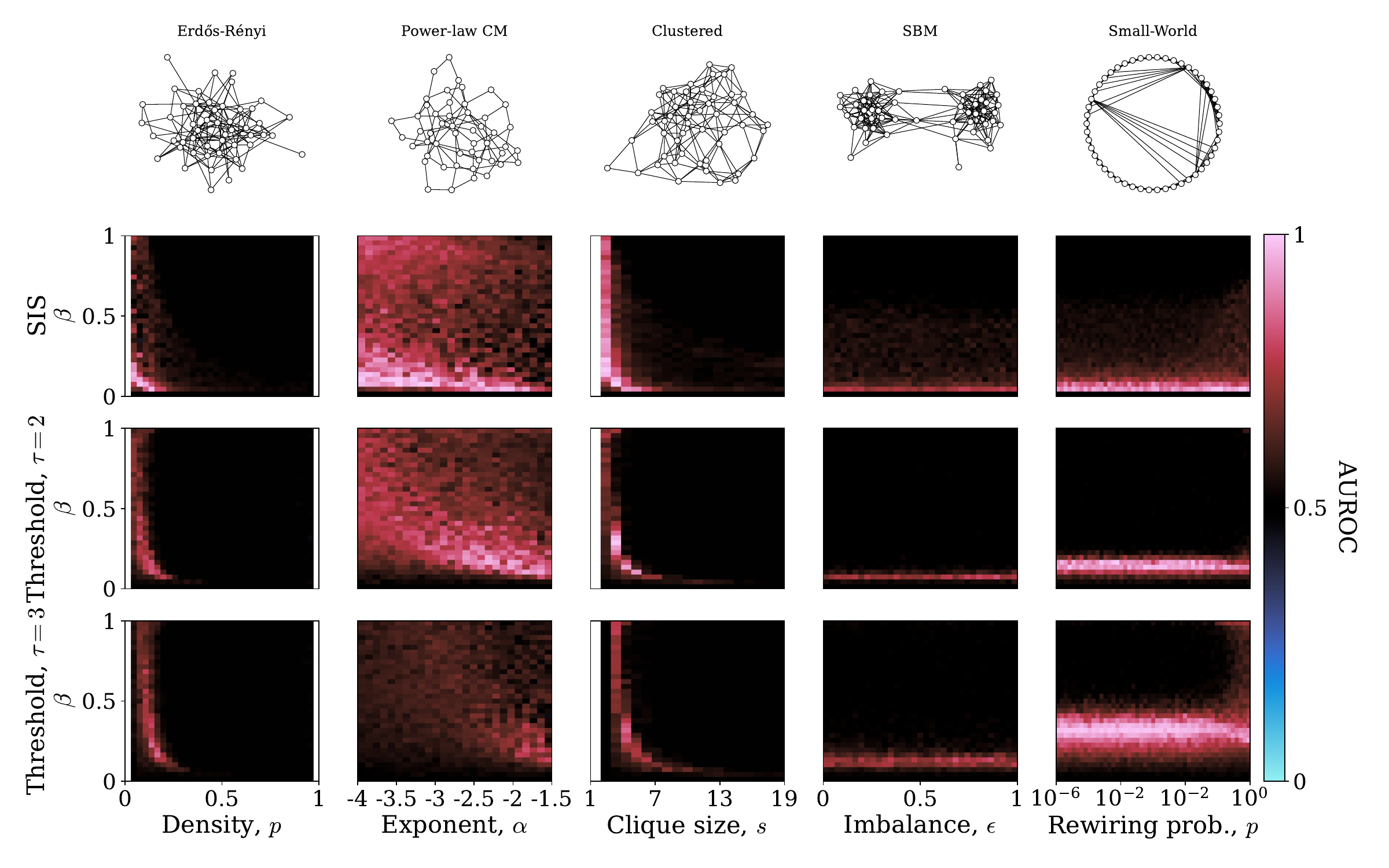}
    \caption{\label{fig:full_absolute_performance}
    The performance of network reconstruction for several common generative network models.
    For each network model, we plot the AUROC with respect to the infectivity and a characteristic structural parameter for three common contagion models: the network SIS model and a variant of the threshold contagion model with two different thresholds, $\tau=2$ and $\tau=3$.
    }
\end{figure*}

Fig.~\ref{fig:full_absolute_performance} plots the AUROC of the SIS model, the threshold model with $\tau=2$, and the threshold model with $\tau=3$ over the set of its dynamical and structural parameters.
This corroborates our results in Figs.~\ref{fig:compare_sis_tau2} and ~\ref{fig:compare_sis_tau3} and provides context for the performance of these algorithms in an absolute sense.
Black regions indicate areas where our algorithm does no better than random, and red regions indicate areas where our algorithm outperforms a random classifier.
The results in this figure illustrate that our main takeaways ---that $R_0$ is sufficient to explain when simple contagion outperforms complex contagion---holds.
For example, for the SBM, the epidemic threshold remains constant in the mean-field limit for a fixed mean degree, and our network reconstruction seems to be independent of the structural parameter, $\epsilon$.

\end{document}